\documentclass[twocolumn,showpacs,preprintnumbers,amsmath,amssymb]{revtex4}

\usepackage{graphicx}
\usepackage{bm}

\input epsf
\begin{document}

\title{On computational irreducibility and the
predictability of complex physical systems}
\author{Navot Israeli}
\author{Nigel Goldenfeld}
\affiliation{Department of Physics, University of Illinois at
Urbana-Champaign, 1110 West Green Street, Urbana, Illinois, 61801-3080.}

\begin{abstract}
Using elementary cellular automata (CA) as an example, we show how to
coarse-grain CA in all classes of Wolfram's classification.  We find
that computationally irreducible (CIR) physical processes can be predictable
and even computationally reducible at a coarse-grained level of
description.  The resulting coarse-grained CA which we construct
emulate the large-scale behavior of the original systems without
accounting for small-scale details. At least one of the CA that can be
coarse-grained is irreducible and known to be a universal Turing
machine.
\end{abstract}

\maketitle

Can one predict the future evolution of a physical process which
is described or modeled by a computationally irreducible (CIR)
mathematical algorithm?  For such systems, in order to know the
system's state after (e.g.) one million time steps, there is no
faster algorithm than to solve the equation of motion a million
time steps into the future. Wolfram has suggested that the
existence of CIR systems in nature is at the root of our apparent
inability to model and understand complex systems \cite{wolf3,wolf5,new_kind_of_science,ilachinski}.

Complex physical systems that are CIR might
therefore seem to be inherently unpredictable.  It is tempting to
conclude from this that the enterprise of physics itself is doomed from
the outset; rather than attempting to construct solvable mathematical
models of physical processes, computational models should be built,
explored and empirically analyzed.  This argument, however, assumes
that infinite precision is required for the prediction of future
evolution.  Usually coarse-grained or even statistical information is
sufficient: indeed, a physical model is usually correct only to a
certain level of resolution, so that there is little interest in
predictions from such a model on a scale outside its regime of
validity.

In this Letter, we report on experiments with nearest neighbour
one-dimensional cellular automata, which show that because in
practice one only seeks coarse-grained information, complex physical
systems can be predictable and even computationally reducible at some
level of description.  The implication of these results is that, at
least for systems whose complexity is the outcome of very simple rules,
useful approximations can be made that enable predictions about future
behavior.

Cellular automata (CA) are dynamical systems composed of a lattice of
cells. Each cell in the lattice can assume a value from a given finite
alphabet. The system evolves in time according to an update rule that
gives a cell's new state as a function of values in its finite
neighborhood. CA were originally introduced by von Neumann and Ulam
\cite{vonneumann} in the 1940's as a possible way of simulating self
reproduction in biological systems. Since then, CA have attracted a
great deal of interest in physics
\cite{wolf1,physicaD10and45,ilachinski,wolf_collected_papers} because they capture
two basic ingredients of many physical systems: 1) they evolve
according to a local uniform rule. 2) CA can exhibit rich behavior even
with very simple update rules. For similar and other reasons, CA have
also attracted attention in computer science
\cite{mitchell98,sarkar00}, biology \cite{eek93}, material science
\cite{raabe02} and other fields.

In early work \cite{wolf2,wolf3,wolf5,new_kind_of_science},
Wolfram proposed that CA can be grouped into four classes of
complexity. Class 1 consists of CA whose dynamics reaches a steady
state regardless of the initial conditions. Class 2 consists of CA
whose long time evolution produces periodic or nested structures. CA
from both of these classes are simple in the sense that their long time
evolution can be deduced from running the system a small number of time
steps. On the other hand, class 3 and class 4 consist of \lq\lq
complex" CA. Class 3 CA produce structures that seem
random. Class 4 CA produce a mixture of random structures and periodic
behavior.
This transition in  CA complexity was later regarded as a
phase transition\cite{langton90} in the CA rule space.
For a review on
CA classification see Refs.\ \onlinecite{mitchell98,ilachinski,new_kind_of_science}. There is no
generally agreed upon algorithm for classifying a given CA. The
assignment of CA to these four classes is somewhat subjective and, we
will argue, may need to be refined.  Based on numerical experiments,
Wolfram hypothesized \cite{wolf2,wolf3,new_kind_of_science} that most
CA from class 3 and 4 are CIR.

There is no unique way to define coarse-graining, but here we will mean
that our information about the CA is {\it locally coarse-grained} in
the sense of being stroboscopic in time, but that nearby cells are
grouped into a supercell according to some specified rule (as is
frequently done in statistical physics).
A system which can be coarse-grained is {\it compactable} since it is
possible to calculate its future time evolution (or some coarse aspects
of it) using a more compact algorithm than its native description. Note
that our use of the term compactable refers to the phase space
reduction associated with coarse-graining, and is agnostic as to
whether or not the coarse-grained system is reducible or irreducible.
Accordingly, we define {\it predictable} to mean that a system is
computationally reducible or has a computationally reducible
coarse-graining. Thus, it is possible to calculate the future time
evolution of a predictable system (or some coarse aspects of it) using
an algorithm which is more compact than both the native and
coarse-grained descriptions.

In order to quantify the implications of looking at coarse-grained
information only, we have systematically attempted to coarse-grain the
256 nearest neighbor one-dimensional binary CA that were the subject of
Wolfram's investigations\cite{new_kind_of_science,wolf1}. The outcome, described in detail below, is
surprising: we found that many CA can be coarse-grained and that in some cases
CIR CA are coarse-grained by computationally
reducible ones. In other words, even though microscopically
a given system might be CIR, its coarse-grained
dynamics can be compactable and predictable.

We start by defining a simple procedure for coarse-graining a CA. Other
constructions are undoubtedly possible.  For simplicity we limit our
treatment to one-dimensional systems with nearest neighbor
interactions. Generalizations to higher dimensions and different
interaction radii are straightforward. Let
$A=\left(a\left(t\right),S_A,f_A\right)$ be a cellular automaton
defined on an array of cells
$a\left(t\right)=\left.\left\{a_n\left(t\right)\right\}
\right|_{n=-\infty}^{\infty}$. Each cell accepts an alphabet of $S_A$
symbols, namely $a_n\left(t\right)\in\left\{0 \dots S_A-1\right\}$. The
values of the cells evolve in time according to the update rule
$a_n\left(t+1\right)=f_A\left[a_{n-1}\left(t\right),a_{n}\left(t\right),
a_{n+1}\left(t\right)\right]$, where
$f_A:\left\{S_A\right\}^3\rightarrow \left\{S_A\right\}$ is the
transition function. The update rule is applied simultaneously to all
the cells and we denote this application by $a\left(t+1\right)=f_A\cdot
a\left(t\right)$.

Our goal is to find a modified CA
$B=\left(b\left(t\right),S_B,f_B\right)$ and an irreversible
coarse-graining
function $b=C\left(a\right)$, which are capable of a coarse-grained
emulation of $A$. For every initial condition $a(0)$, $B$ and $C$ must
satisfy:
\begin{equation}
C\left(f_A^{T\cdot t} \cdot a(0)\right)=f_B^t\cdot C\left(a(0)\right).
\label{coarse-graining_def}
\end{equation}
Namely, running the original CA for $T\cdot t$ time steps and then
coarse-graining is equivalent to coarse-graining the initial condition
and then running the modified CA $t$ time steps. The constant $T$ is a
time scale associated with the coarse-graining.

To search for explicit coarse-graining rules, we define the $N$'th
block version $A^N=\left(a^N,S_{A^N},f_{A^N}\right)$ of $A$.
$S_{A^N}=\left(S_A\right)^N$ and each cell in $A^N$ represents a block
of $N$ cells in $A$. Cell values are translated between $A$ and $A^N$
according to the base $S_A$ value of $N$ cells in $A$. The transition
function $f_{A^N}:\left\{S_{A^N}\right\}^3\rightarrow
\left\{S_{A^N}\right\}$ is computed by running $A$ for $N$ time steps on
all possible initial conditions of length $3N$. In this way $A^N$
computes in one time step $N$ time steps of $A$. Note that
$A^N$ is not a coarse-graining of $A$ because no information was lost
in the cell translation.

Next we attempt to generate the coarse CA $B$ by projecting the
alphabet of $A^N$ on a subset of $\left\{0\dots S_{A^N}-1\right\}$.
This is the key step where information is being lost, a manipulation
which distinguishes between coarse-graining and emulation
blocking transformations \cite{new_kind_of_science,ilachinski}.
The transition function $f_B$
is constructed from $f_{A^N}$ by projecting its arguments and
outcome:
\begin{equation}
f_B\left[\overline{x_1},\overline{x_2},\overline{x_3}\right]=
\overline{f_{A^N}\left[\overline{x_1},\overline{x_2},\overline{x_3}
\right]}.
\end{equation}
Here $\overline{x}\equiv P\left(x\right)$ denotes the projection operation. This
construction is possible only if
\begin{equation}
\overline{f_{A^N}\left[x_1,x_2,x_3\right]}=
\overline{f_{A^N}\left[y_1,y_2,y_3\right]},\;\;
\forall \left(x,y|\overline{x_i}=\overline{y_i}\right).
\label{trans_func_projection}
\end{equation}
Otherwise, $f_B$ is multi-valued and our
coarse-graining attempt fails for the specific choice of $N$ and $P$.

In cases where the above conditions are satisfied, the resulting CA $B$
is a coarse-graining of $A^N$ with a time scale $T=1$. For every step
$a^N_n(t+1)=f_{A^N}\left[a^N_{n-1}(t),a^N_n(t),a^N_{n+1}(t)\right]$
of $A^N$, $B$ makes the move
\begin{eqnarray}
b_n(t+1)&=&f_B\left[b_{n-1}(t),b_n(t),b_{n+1}(t)\right] \\
&=&\overline{f_{A^N}\left[\overline{a^N_{n-1}(t)},
\overline{a^N_n(t)},\overline{a^N_{n+1}(t)}\right]}=\overline{a^N_n(t+1)}\;,
\nonumber
\end{eqnarray}
where we have used Eq.\ (\ref{trans_func_projection}) in the last step.
$B$ therefore satisfies Eq.\ (\ref{coarse-graining_def}) with $P$
as the coarse-graining function. Since a single time step of $A^N$
computes $N$ time steps of $A$, $B$ is also a coarse-graining of
$A$ with a coarse-grained time scale $T=N$. The coarse-graining
function $C$ in this case is composed of the translation from $A$
to $A^N$ followed by the projection operator $P$.  Analogies of
these operators have been used in attempts to reduce the
computational complexity of certain stochastic partial
differential equations \cite{hou01,degenhard02}. Similar ideas
have been used to calculate critical exponents in probabilistic CA
\cite{oliveira97,monetti98}.

It is interesting to notice that the above coarse-graining procedure
can lose two very different types of dynamic information. To see this,
consider Eq.\ (\ref{trans_func_projection}). This equation can be
satisfied in two ways. In the first case
\begin{equation}
f_{A^N}\left[x_1,x_2,x_3\right]=f_{A^N}\left[y_1,
y_2,y_3\right],\;\;
\forall \left(x,y|\overline{x_i}=\overline{y_i}\right)\;,
\label{irrelevant_cond}
\end{equation}
which necessarily leads to Eq.\  (\ref{trans_func_projection}).
$f_{A^N}$ in this case is insensitive to the projection of its
arguments. The distinction between two variables which are identical
under projection is therefore {\it irrelevant} to the dynamics of
$A^N$, and by construction to the long time dynamics of $A$. By
eliminating irrelevant degrees of freedom (DOF), coarse-graining of
this type removes information which is redundant on the microscopic
scale. The coarse CA in this case accounts for all possible long time
trajectories of the original CA and the complexity classification of
the two CA is therefore the same.

In the second case Eq.\ (\ref{trans_func_projection}) is satisfied even
though Eq.\ (\ref{irrelevant_cond}) is violated.
Here the distinction between two variables which are identical under
projection is {\it relevant} to the dynamics of $A$. Replacing $x$ by
$y$ in the initial condition may give rise to a difference in the
dynamics of $A$. Moreover, the difference can be (and in many occasions
is) unbounded in space and time. Coarse-graining in this case is
possible because the difference is constrained in the symbol space by
the projection operator. Namely, projection of all such different
dynamics results in the same coarse-grained behavior. Note that the
coarse CA in this case cannot account for all possible long time
trajectories of the original one. It is therefore possible for the
original and coarse CA to fall into different complexity
classifications.

Coarse-graining by elimination of relevant DOF removes information
which is not redundant with respect to the original system. The
information becomes redundant only when moving to the coarse scale. In
fact, \lq\lq redundant" becomes a subjective qualifier here since it
depends on our choice of coarse description. In other words, it depends
on what aspects of the microscopic dynamics we want the coarse CA to
capture. In a sense, this is analogous to the subtleties encountered in
constructing renormalization group transformations for the critical
behavior of antiferromagnets \cite{nigelbookp268,leeuwen75}.

\begin{figure}
\centerline{
\epsfxsize=42mm \epsffile{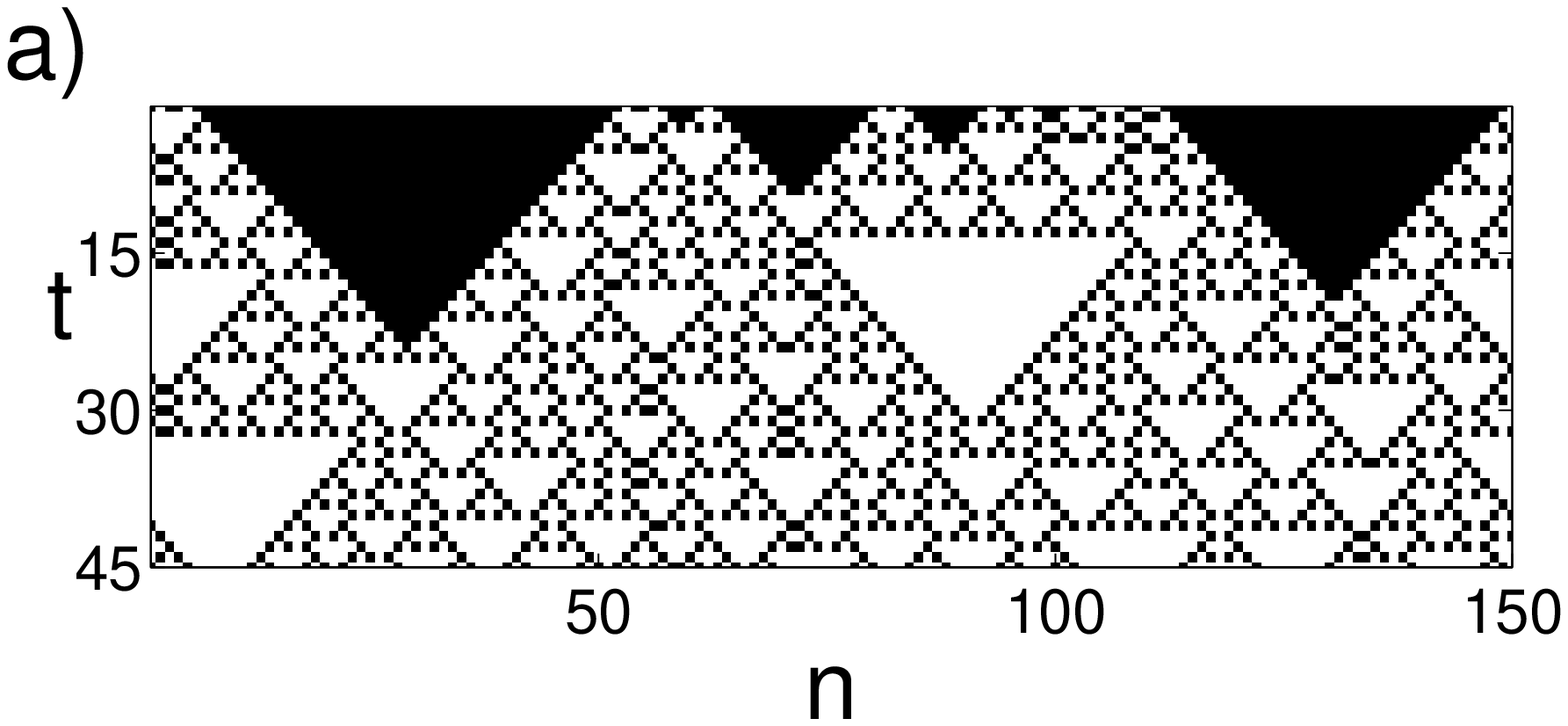}
\hspace{0.5mm}
\epsfxsize=42mm \epsffile{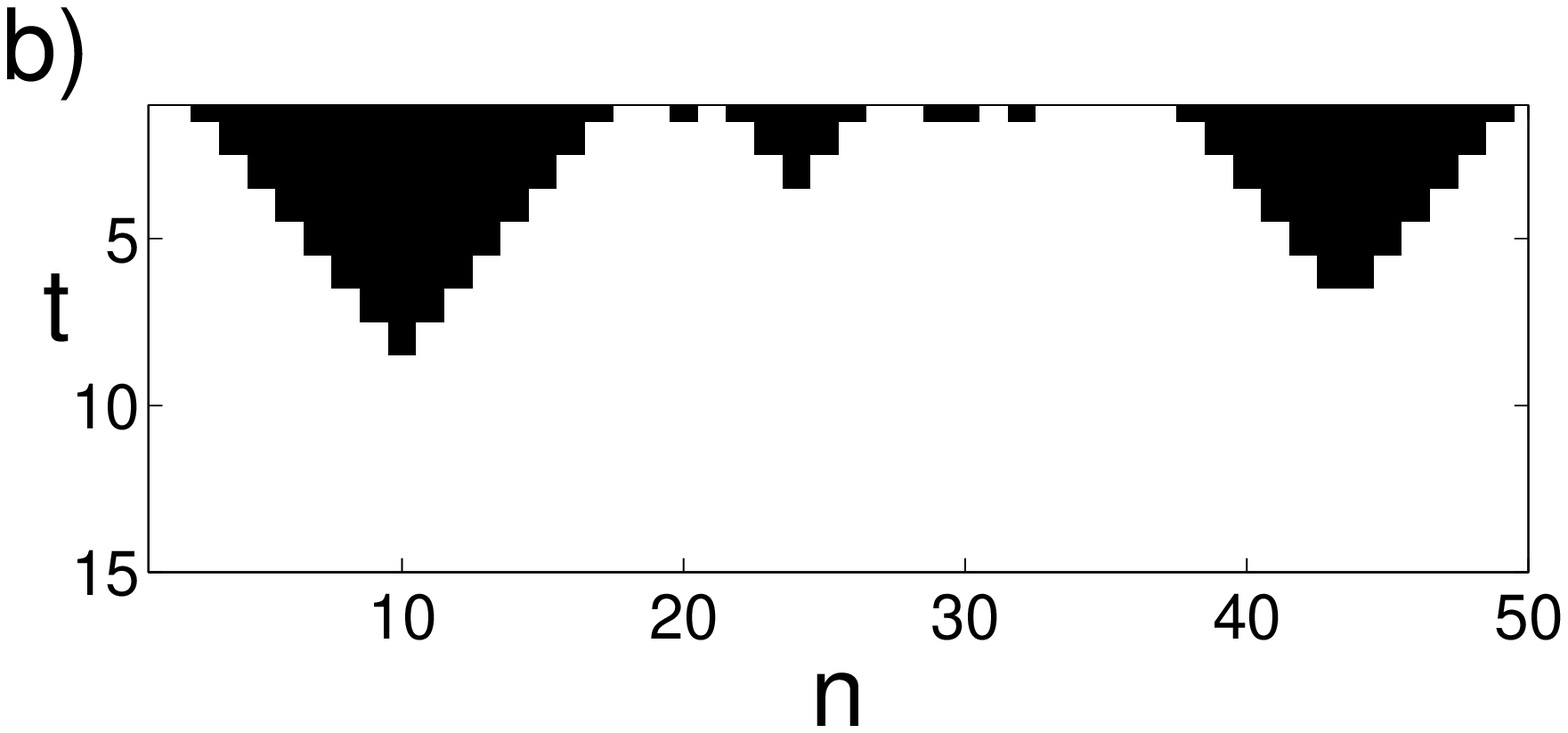}}
\centerline{
\epsfxsize=42mm \epsffile{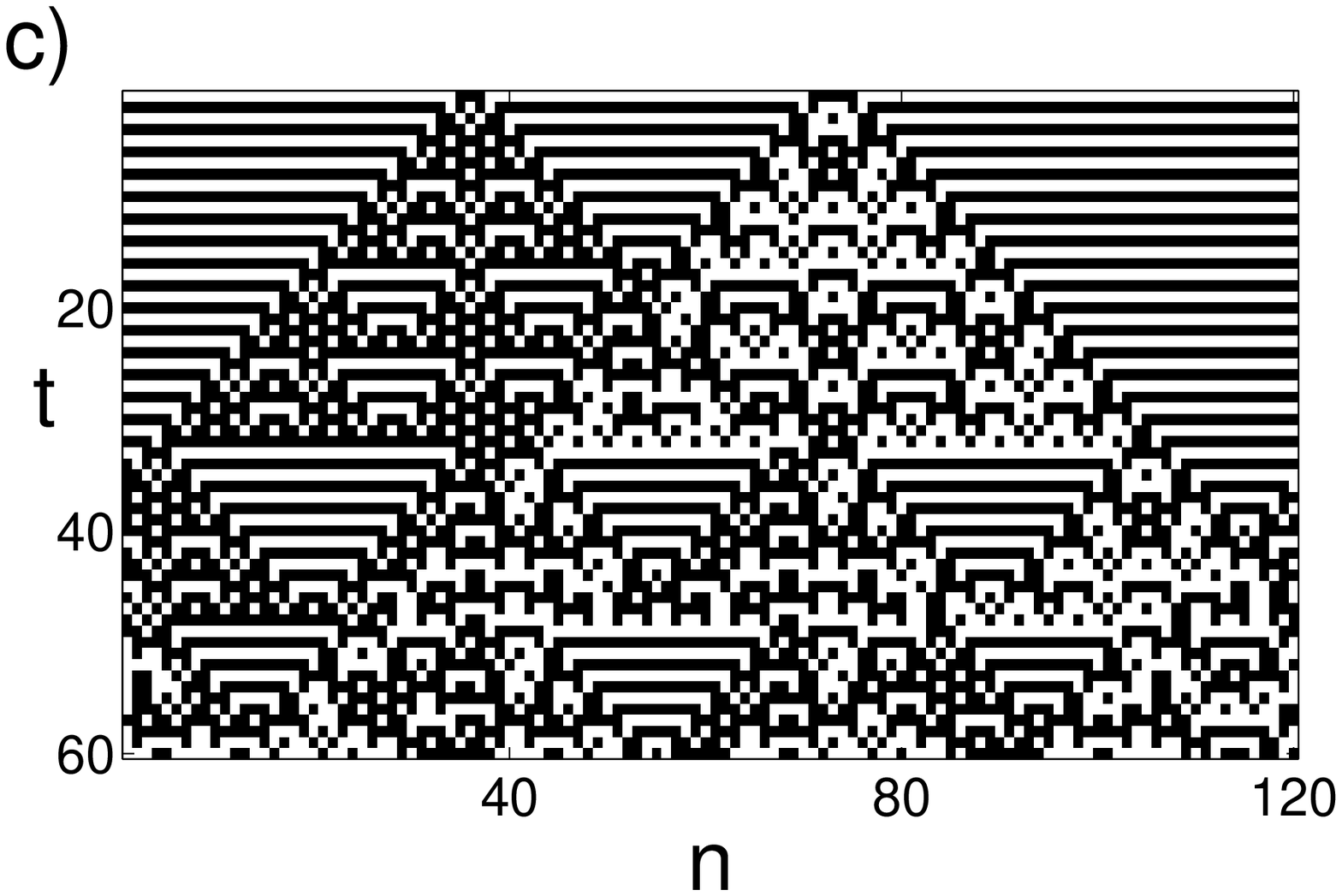}
\hspace{0.5mm}
\epsfxsize=42mm \epsffile{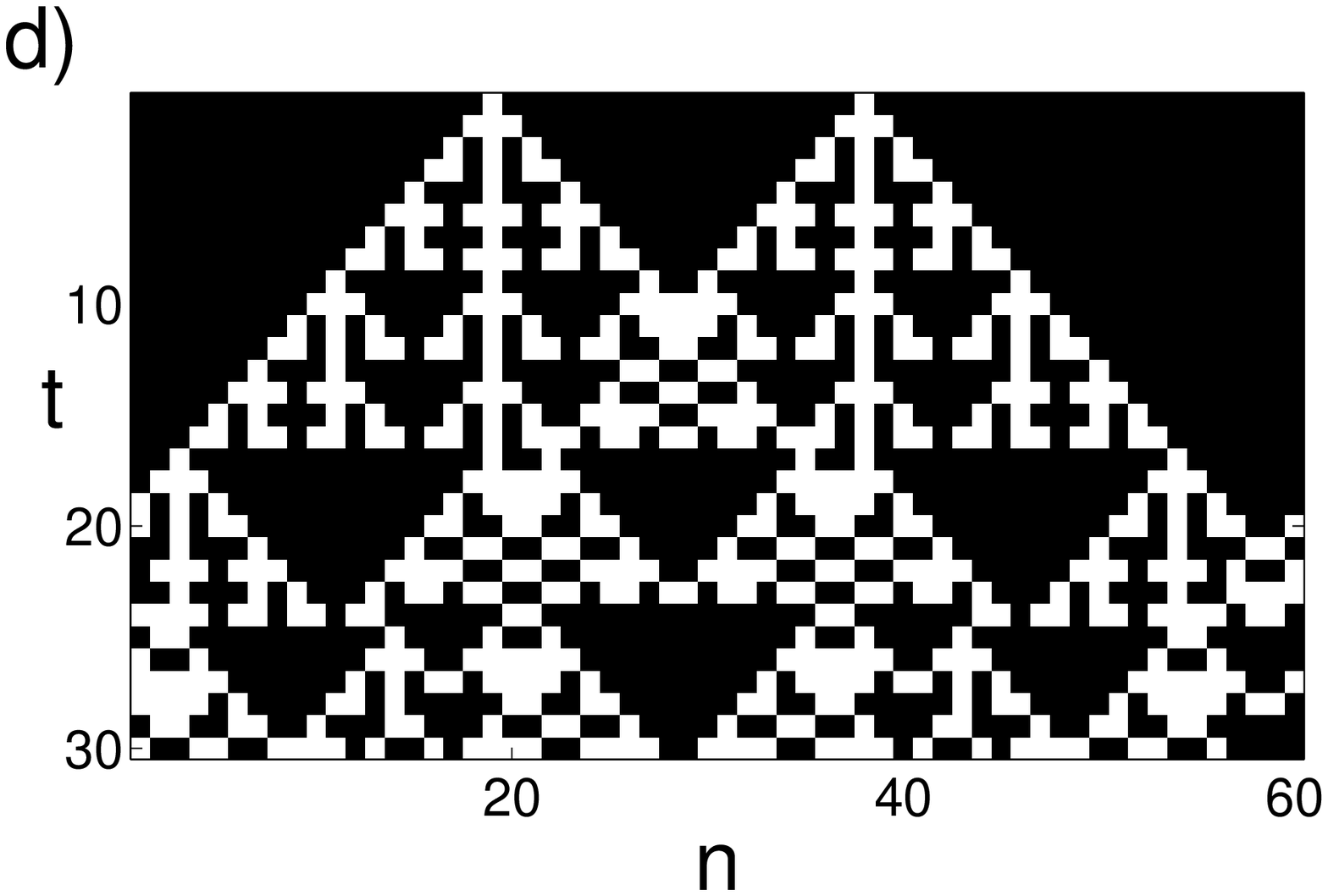}}
\caption{Examples of coarse-graining transitions. (a) and (b) show coarse-graining rule 146 by rule 128. (a) shows results of running rule 146. The top line is the initial condition and time progress from top to bottom. (b) shows the results of running rule 128 with the coarse grained initial condition from (a). (c) and (d) show coarse-graining rule 105 by rule 150. (c) shows rule 105 and (d) shows rule 150.}
\label{example_fig}
\end{figure}

We now give specific examples of coarse-graining CA. In the sequel, CA
rules are numbered using Wolfram's notation\cite{wolf1,new_kind_of_science}.
Figure \ref{example_fig} (a) and (b) shows a coarse-graining of rule 146 by rule 128.
Rule 146 produces a complex, seemingly random behavior which falls into the class 3 group.
We use a super-cell size $N=3$, and project the alphabet $\{0\dots7\}$
of the super-cells back to the $\{0,1\}$ alphabet with $P(7)=1$ and
$P(x\neq7)=0$\cite{projection_footnote}.
A triplet of cells in rule 146 are therefore
coarse-grained to a single cell and the value of the coarse cell is $1$
only when the triplet are all $1$. Using this projection operator we
construct the transition function of the coarse CA which is found to be
rule 128, a class 1 CA.  This choice of coarse-graining
eliminates the small scale details of rule 146. Only structures of
lateral size of three or more cells are accounted for. The decay of
such structure in rule 146 is accurately described by rule 128.

Note that a class 3 CA was coarse-grained to a class 1 CA in the above
example. Our gain was therefore two-fold. In addition to the phase space
reduction associated with coarse-graining we have also achieved a
reduction in complexity. Our procedure found
predictable
coarse-grained aspects of the dynamics even though the small scale
behavior of rule 146 is complex, potentially CIR.
As we explained earlier, this type of simplification can be achieved only by
eliminating relevant DOF.

As a second example we show a transition
between rules with a comparable complexity. Fig.\ \ref{example_fig} (c) and (d) shows
a coarse-graining of rule 105 by rule 150. $N=2$ in this example and $P(x)=1$ only
when $x=0,3$\cite{projection_footnote}.

The coarse-graining procedure we described above is not constructive,
but instead is a self-consistency condition on a putative
coarse-graining rule with a specific block size $N$ and projection
operator $P$. In many cases the coarse-graining fails and one must try
other choices of $N$ and $P$. Can all CA be coarse-grained? If not,
which CA can be coarse-grained and which cannot?

\begin{figure}[ht]
\epsfxsize=\columnwidth
\begin{center}
\leavevmode
\epsffile{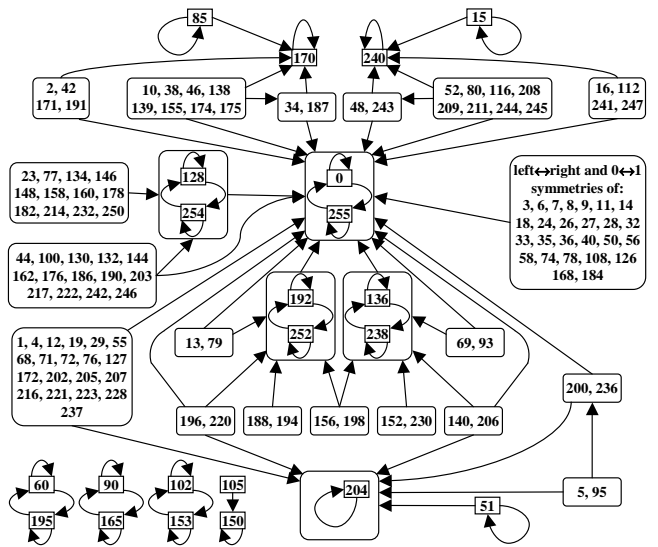}
\end{center}
\vspace{-4mm}
\caption{
Coarse-graining transitions within the family of 256
elementary CA. Only transitions with a cell block size $N=2,3,4$ are
shown. An arrow indicates that the origin rules can be
coarse-grained by the target rules and may correspond to several choices of $N$ and $P$.
} \label{mapfigure}
\end{figure}

To answer these questions we tried systematically to coarse-grain
Wolfram's 256 elementary rules. We applied the coarse-graining
procedure to each rule and scaned the $N$,$P$ space for valid solutions. In this
way we were able to coarse-grain 240 out of the 256 CA \cite{different_projections_footnote}. These 240
coarsen-able rules include members of all four classes. Many elementary
CA can be coarse-grained by other elementary CA. Figure
\ref{mapfigure} shows a map of the coarse-graining transitions that we
found within the family of elementary rules. Only coarse-grainings
with $N\leq 4$ are shown due to limited computing power. Other
transitions may exist with larger $N$.
We observe that rule complexity never increases along the map's
transitions, i.e., coarse-graining introduces (at least here)
partial order among CA rules.

As mentioned above, we were unable to coarse-grain 16
elementary rules. 12 out of the 16 are the class 3 rules
30,45,106 and their symmetries. The other four
are the class 2 rule 154 and it's symmetries. We don't
know if our inability to coarse-grain these 16 rules
comes from limited computing power or from something deeper.

It is worth noticing that a subset of the fixed points in the
transition map is composed of all elementary additive rules (see page
952 in Ref.\ \onlinecite{new_kind_of_science}) and their symmetries.
This result is not limited to elementary rules. All additive CA whose
alphabet sizes $S$ are prime numbers coarse-grain themselves with $N=S$ and $P$ the modulo $S$ sum of $S$ cells. We conjecture
that there are situations where reducible fixed points exist for a wide
range of systems, analogous to the emergence of amplitude equations in
the vicinity of bifurcations points.

Coarse-graining transitions can also exit the elementary CA
family. This happens whenever the alphabet of the coarse CA
consists of more than two symbols. One such example which is of
special importance is rule 110. Rule 110 is interesting because
this class 4 CA is universal \cite{new_kind_of_science} in the
Turing sense \cite{herken} and is therefore CIR \cite{wolf2}. It
is capable of emulating all computations done by other computing
devices in general and CA in particular.

We found several ways to coarse-grain rule 110. Using $N=6$, it is
possible to project the 64 possible states onto an alphabet of 63
symbols. A more impressive reduction in the
alphabet size is
obtained by going to larger values of $N$.
For $N=7,8,9,10,11$ we found an alphabet reduction of $6/128$, $22/256$, $67/512$,
$181/1024$ and $463/2048$ respectively.
We expect this behavior to
persist for larger values of $N$.

Another interesting coarse-graining of rule 110 that we found is the
transition to rule 0. Rule 0 has the trivial dynamics where all initial
states evolve to the null configuration in a single time step. The transition to rule 0 is possible because the cell combination \lq\lq01010" is not 
generated by rule 110 and can only appear in the initial state. Coarse-graining by rule 0 is achieved using
$N=5$ and projecting \lq\lq01010" to 1 and all
other five cell combinations to 0. This example is important because it
shows that even though rule 110 is CIR it has a
predictable coarse-grained dynamics (however trivial).
To our knowledge rule 110 is the only proven CIR elementary CA and
therefore this is the only example of irreducible to reducible transition
between elementary rules that we found.

We did find other complex, {\it potentially} CIR
rules that can be coarse-grained by reducible CA.
Rules 18, 54, 126 and their symmetries are coarse-grained by rule 0.
As we showed above, rule 146 and it's symmetries can be coarse-grained by
rule 128 in a non-trivial way. We don't know if these rules are CIR
for lack of proof. Nevertheless, non-trivial irreducible to reducible
transitions can in principle exist.
Consider for example the CIR CA generated by the product of rules 110 and 128.
Let the CA alphabet be $\{a,b\}\times \{0,1\}$, where the letters evolve according
to rule 110 and the digits according to 128. We can recover the reducible,
coarse-grain-able rule 128 by projecting the alphabet onto $\{0,1\}$.

The fact that CIR rules can be coarse-grained and
that they has predictable coarse-grained dynamics shows that CIR
is not a good measure of physical
complexity. As in the case of rule 110, a CIR
system may still yield an efficient predictable theory, provided that
we are willing to ask coarse-grained questions. It seems that a better
classification of physical complexity is related to what classes of
projection operator are required to coarse-grain the system: local,
real space projections or more complex non-geometric projections?

In summary, we have found that many CA, including CIR
ones can be locally coarse-grained in space and time. In
some cases CIR systems are predictable, if
coarse-grained information only is required.

NG wishes to thank Stephen Wolfram for useful discussions and correspondence.
This work was partially supported by the National Science Foundation
through grant NSF-DMR-99-70690 (NG) and by the National Aeronautics
and Space Administration through grant NAG8-1657.


\begin{thebibliography}{23}
\bibitem[{pro()}]{projection_footnote}
\bibinfo{note}{$P^2\neq P$ because symbols were re-labeled for convenience.}
\bibitem[{dif()}]{different_projections_footnote}
\bibinfo{note}{For a given $N$ we often found several possible $P$'s, capturing
  different coarse-grained aspects of the origin CA.}
\end{thebibliography}

\end{document}